# A Convex Approximation of the Relaxed Binaural Beamforming Optimization Problem

Andreas I. Koutrouvelis, Richard C. Hendriks, Richard Heusdens and Jesper Jensen

*Abstract*—**The recently proposed relaxed binaural beamforming (RBB) optimization problem provides a flexible trade-off between noise suppression and binaural-cue preservation of the sound sources in the acoustic scene. It minimizes the output noise power, under the constraints which guarantee that the target remains unchanged after processing and the binaural-cue distortions of the acoustic sources will be less than a user-defined threshold. However, the RBB problem is a computationally demanding non-convex optimization problem. The only existing suboptimal method which approximately solves the RBB is a successive convex optimization (SCO) method which, typically, requires to solve multiple convex optimization problems per frequency bin, in order to converge. Convergence is achieved when all constraints of the RBB optimization problem are satisfied. In this paper, we propose a semi-definite convex relaxation (SDCR) of the RBB optimization problem. The proposed suboptimal SDCR method solves a single convex optimization problem per frequency bin, resulting in a much lower computational complexity than the SCO method. Unlike the SCO method, the SDCR method does not guarantee user-controlled upper-bounded binaural-cue distortions. To tackle this problem we also propose a suboptimal hybrid method which combines the SDCR and SCO methods. Instrumental measures combined with a listening test show that the SDCR and hybrid methods achieve significantly lower computational complexity than the SCO method, and in most cases better trade-off between predicted intelligibility and binaural-cue preservation than the SCO method.**

## I. Introduction

Binaural beamforming (see e.g., [1] for an overview), also known as binaural spatial filtering, plays an important role in binaural hearing-aid (HA) systems [2]. Binaural beamforming is typically described as an optimization problem, where the objective is to i) minimize the output noise power, ii) preserve the target sound source at the left and right HA reference microphone, and iii) preserve the binaural cues of all sound sources after processing. The microphone array, which is typically mounted on the HA devices, has only a few microphones and, thus, there is only limited freedom (i.e., a small feasibility set) to search for a good compromise between the three aforementioned goals. Besides the challenge in finding a good trade-off among all these goals, the complexity should remain as low as possible, due to the limited computational power of the HA devices.

The binaural minimum variance distortionless response (BMVDR) BF [1] provides the maximum possible noise suppression among all binaural target-distortionless BFs [3]. Unfortunately, the BMVDR severely distorts the binaural-cues of the residual noise at the output of the filter. Specifically, the residual noise inherits the intaraural transfer function of the target and, hence, sounds as originating from the target's direction [1]. The lack of spatial separation between the target and the noise after processing, may not only provide an unnatural impression to the user, but may also negatively effect the intelligiblity [4]. In [5], [6], the BMVDR was compared with an oracle-based (i.e., non-practically implementable) method in several noise fields. The oracle-based method has the same noise suppression as the BMVDR, but does not introduce any binaural-cue distortions at the output. The spatially correct oracle-based method achieved an improvement of about 3 dB in SRT-50[1] over the BMVDR. Therefore, there are several reasons to seek for methods that simultaneously provide the maximum possible noise suppression and binaural-cue preservation of all sources in the acoustic scene.

Several modifications of the BMVDR BF have been proposed, which can be roughly categorized into two groups. The first group consists of BFs that add or maintain a portion of the unprocessed scene at the output of the filter (see e.g., [5], [7]–[9]). The second group consists of BFs, whose optimization problems have the same objective function as the BMVDR, but introduce extra equality [3], [10], [11] or inequality [12] constraints in order to preserve the binaural cues of the interferers after processing. Such additional constraints in the optimization problem results in less degrees of freedom for noise reduction. With equality constraints, closed-form solutions may be derived, but the degrees of freedom can be easily exhausted when multiple interferers exist in the acoustic scene, resulting in poor noise reduction. On the other hand, inequality constraints provide more flexibility and can approximately preserve the binaural cues of, typically, many more acoustic sources, or for the same number of acoustic sources provide larger amount of noise reduction [12]. Unfortunately, closed-form solutions do not exist for the inequality-constrained binaural BFs and, thus, iterative methods with a larger complexity are used instead.

Recently, the relaxed binaural beamforming (RBB) optimization problem was proposed, which uses inequality constraints to preserve the binaural cues of the interfering sources [12]. The inequality constraints in the RBB are not convex, resulting in a non-convex optimization problem. In [12], a suboptimal successive convex optimization (SCO) method was proposed to approximately solve the RBB prob-

This work was supported by the Oticon Foundation and NWO, the Dutch Organisation for Scientific Research.

[1]Speech reception threshold (SRT)-50 is the SNR in which a 50% correct recognition of words is achieved.

lem. In most cases, the SCO method needs to solve more than one convex optimization problem, per frequency bin, in order to converge. Convergence is achieved when all constraints of the RBB problem are satisfied. As a result, the SCO method guarantees an upper-bounded binaural-cue distortion of the interferers (as expressed by the interaural transfer function error), where the upper bound is controlled by the user.

Unfortunately, the SCO method is computationally very demanding due to its need to solve multiple convex optimization problems, per frequency bin, in order to converge. In this paper, we propose a semi-definite convex relaxation (SDCR) of the RBB optimization problem, which is significantly faster than the SCO method. This is because, the SDCR method requires to solve only one convex optimization problem per frequency bin. The main drawback of the SDCR method is that it does not guarantee user-controlled upper-bounded binaural-cue distortions as the SCO method. We solve this issue by combining the SDCR and SCO methods into a sub-optimal hybrid method. The hybrid method guarantees user-controlled upper-bounded binaural-cue distortions, and still has a significantly lower computational complexity than the SCO method. Simulation experiments combined with listening tests show that both proposed methods, in most cases, provide a better trade-off between noise reduction and binaural-cue preservation than the SCO method.

## II. Signal Model and Notation

We assume that there is one target point-source signal, $r$ point-source interferers, additive diffuse noise, and two HAs with $M$ microphones in total. The processing is accomplished per time-frequency bin independently. Neglecting time-frequency indices for brevity, the acquired $M$-element noisy vector in the DFT domain, for a single time-frequency bin, is given by

$$\mathbf{y} = \underbrace{s\mathbf{a}}_{\mathbf{x}} + \underbrace{\sum_{i=1}^{r} v_i \mathbf{b}_i + \mathbf{u}}_{\mathbf{n}} \in \mathbb{C}^{M\times 1}, \tag{1}$$

where $s$ and $v_i$ are the target and $i$-th interferer signals at the original locations; $\mathbf{a}$ and $\mathbf{b}_i$ the acoustic transfer function (ATF) vectors of the target and $i$-th interferer, respectively; $\mathbf{u}$ the diffuse background noise, and $\mathbf{n}$ the total additive noise.

Assuming statistical independence between all sources, the noisy cross-power spectral density matrix is given by

$$\mathbf{P_y} = \mathrm{E}[\mathbf{y}\mathbf{y}^H] = \mathbf{P_x} + \mathbf{P_n} \in \mathbb{C}^{M\times M}, \tag{2}$$

with $\mathbf{P_x} = \mathrm{E}[\mathbf{x}\mathbf{x}^H] = p_s\mathbf{a}\mathbf{a}^H$ and $\mathbf{P_n} = \mathrm{E}[\mathbf{n}\mathbf{n}^H]$ the target and noise cross-power spectral density matrices, respectively, and $p_s = \mathrm{E}[|s|^2]$ the power spectral density of the target signal.

## III. Binaural Beamforming Preliminaries

Binaural BFs consist of two spatial filters, $\mathbf{w}_L$, $\mathbf{w}_R \in \mathbb{C}^{M\times 1}$, which are both applied to the noisy measurements producing two different outputs given by

$$\begin{bmatrix} \hat{x}_L \\ \hat{x}_R \end{bmatrix} = \begin{bmatrix} \mathbf{w}_L^H\mathbf{y} \\ \mathbf{w}_R^H\mathbf{y} \end{bmatrix}, \tag{3}$$

where $\hat{x}_L, \hat{x}_R$ are played back by the loudspeakers of the left and right HAs, respectively. Note that the subscripts $L$ and $R$ are also used to refer to the two elements of the vectors in Eq. (1) associated with the left and right reference microphones of the binaural BF. Here, we select the first and the $M$-th microphones as reference microphones and, thus, $y_L = y_1$ and $y_R = y_M$. The same applies to all the other vectors in Eq. (1).

All BFs considered in this paper are target-distortionless. Their goal is not only noise supression, but also preservation of the binaural cues of all sources in the acoustic scene. In this paper, we mainly focus on preserving, after processing, the perceived location of the point sources. A simple way of measuring the binaural cues of a point source is via the interaural transfer function (ITF), which is a function of the ATF vector of the source [13]. The ITF of the $i$-th interferer before and after applying the spatial filter is given by [13]

$$\mathrm{ITF}_i^{\text{in}} = \frac{b_{iL}}{b_{iR}}, \quad \mathrm{ITF}_i^{\text{out}} = \frac{\mathbf{w}_L^H\mathbf{b}_i}{\mathbf{w}_R^H\mathbf{b}_i}. \tag{4}$$

The input and output ITF of the target is expressed similarly. Ideally, to preserve all spatial cues of the point sources, a binaural BF will produce the same ITF output as the input for all point sources. In practice, this is very difficult to achieve, when the number of interferers, $r$, is large and the number of microphones, $M$, is small [12]. As a result, most BFs will introduce some distortion to the ITF output, resulting in a non-zero ITF error given by [12]

$$\mathrm{ITF}_i^{\text{e}} = \left|\mathrm{ITF}_i^{\text{out}} - \mathrm{ITF}_i^{\text{in}}\right| = \left|\frac{\mathbf{w}_L^H\mathbf{b}_i}{\mathbf{w}_R^H\mathbf{b}_i} - \frac{b_{iL}}{b_{iR}}\right| \geq 0. \tag{5}$$

### A. BMVDR Beamforming

The BMVDR BF [1] achieves the maximum possible noise suppression among all binaural BFs and is obtained from the following simple optimization problem [1], [3]:

$$\hat{\mathbf{w}}_L, \hat{\mathbf{w}}_R = \underset{\mathbf{w}_L,\mathbf{w}_R}{\arg\min} \begin{bmatrix}\mathbf{w}_L^H & \mathbf{w}_R^H\end{bmatrix} \tilde{\mathbf{P}} \begin{bmatrix}\mathbf{w}_L \\ \mathbf{w}_R\end{bmatrix}$$
$$\text{s.t.} \quad \mathbf{w}_L^H\mathbf{a} = a_L^* \quad \mathbf{w}_R^H\mathbf{a} = a_R^*, \tag{6}$$

where

$$\tilde{\mathbf{P}} = \begin{bmatrix}\mathbf{P_n} & \mathbf{0} \\ \mathbf{0} & \mathbf{P_n}\end{bmatrix}. \tag{7}$$

The optimization problem in Eq. (6) provides closed-form solutions to the left and right spatial filters given by [1], [3]

$$\hat{\mathbf{w}}_L = \frac{\mathbf{P_n}^{-1}\mathbf{a}a_L^*}{\mathbf{a}^H\mathbf{P_n}^{-1}\mathbf{a}}, \quad \hat{\mathbf{w}}_R = \frac{\mathbf{P_n}^{-1}\mathbf{a}a_R^*}{\mathbf{a}^H\mathbf{P_n}^{-1}\mathbf{a}}. \tag{8}$$

It can easily be shown, that the output ITF of the $i$-th interferer of the BMVDR spatial filter is given by [3], [12]

$$\mathrm{ITF}_i^{\text{out}} = \frac{a_L}{a_R}, \tag{9}$$

which is the ITF input of the target. Therefore, all interferers sound as coming from the target direction after applying the BMVDR spatial filter. The BMVDR ITF error of the $i$-th interferer is given by [12]

$$\mathrm{ITF}_i^{\text{e,BMVDR}} = \left|\frac{a_L}{a_R} - \frac{b_{iL}}{b_{iR}}\right|. \tag{10}$$

### *B. Relaxed Binaural Beamforming*

The relaxed binaural beamforming (RBB) optimization problem, introduced in [12], uses additional inequality constraints (compared to the BMVDR problem) to preserve the binaural cues of the interferers. The RBB problem is given by [12]

$$\begin{aligned}\hat{\mathbf{w}}_L, \hat{\mathbf{w}}_R = \arg\min_{\mathbf{w}_L,\mathbf{w}_R} & \begin{bmatrix}\mathbf{w}_L^H & \mathbf{w}_R^H\end{bmatrix} \tilde{\mathbf{P}} \begin{bmatrix}\mathbf{w}_L \\ \mathbf{w}_R\end{bmatrix} \\ \text{s.t.} \quad & \mathbf{w}_L^H \mathbf{a} = a_L^* \quad \mathbf{w}_R^H \mathbf{a} = a_R^*, \\ & \left| \frac{\mathbf{w}_L^H \mathbf{b}_i}{\mathbf{w}_R^H \mathbf{b}_i} - \frac{b_{iL}}{b_{iR}} \right| \leq \mathcal{E}_i, \; i = 1, \cdots, m \leq r,\end{aligned} \tag{11}$$

where

$$\mathcal{E}_i = c_i \text{ITF}_i^{\text{e,BMVDR}}, \quad 0 \leq c_i \leq 1.$$

Note that $\mathcal{E}_i$ is $c_i$ times the ITF error of the $i$-th interferer of the BMVDR BF [12]. Recall that the BMVDR causes full collapse of the binaural cues of the interferers towards the binaural cues of the target. Therefore, the inequality constraints in Eq. (11) control the percentage of collapse. A small $c_i$ implies good preservation of binaural cues of the $i$-th interferer, but a smaller feasibility set and, thus, less noise reduction. On the other hand, a large $c_i$ implies worse binaural-cue preservation, but more noise reduction.

It is clear from the above that the additional inequality constraints of the RBB problem require the knowledge of the (relative) ATF vectors of the interferers. In practice, interferers' (R)ATF vectors are unknown and estimation is required. Several methods for estimating RATF vectors exist (see e.g., [14] for an overview). An alternative approach is to use pre-determined ancechoic (R)ATF vectors of fixed azimuths around the head of the user, as proposed in [15]. These pre-determined (R)ATF vectors are acoustic scene independent and need to be obtained once for each user. This is useful when the (R)ATF vectors of the interferers are difficult to estimate, because e.g., the locations of the interferers relative to the head of the user are non-static. It is worth noting that by using pre-determined (R)ATF vectors, a larger number of inequality constraints, $m > r$, is typically used in Eq. (11). This is because we do not know where the interferers are located and we would like to cover the entire space around the head of the user.

If $c_i > 0, i = 1, \cdots, m$, the inequality constraints of the optimization problem in Eq. (11) are non-convex. As a result, the optimization problem in Eq. (11) is non-convex. In [12], a suboptimal successive convex optimization (SCO) method [12], described in Section III-C, was proposed to approximately solve the RBB problem.

### *C. Successive Convex Optimization method*

The successive convex optimization (SCO) method [12] approximately solves the RBB problem by solving multiple convex optimization problems per frequency bin. The SCO method converges, when all constraints of the RBB problem in Eq. (11) are satisfied. It has been shown that the SCO method always converges to a solution satisfying the constraints of the RBB problem if $m \leq 2M - 3$. This means that if the (R)ATF vectors of the interferers have been estimated accurately enough, the SCO method will guarantee user-controlled upper-bounded ITF error of the interferers [12]. For $m > 2M - 3$, no guarantees exist for convergence. In case the method does not converge, it stops after solving a pre-defined maximum number of convex optimization problems, $k_{\text{max}}$. Nevertheless, for a reasonable number of inequality constraints, $m$, it has been experimentally shown that the SCO method always converges [12], [15]. A disadvantage of the SCO method is that it has been experimentally shown in [12], that for larger $c_i$ values, the SCO method converges to solutions further away from the boundary of the inequality constraints of the RBB problem. This results in a better binaural-cue preservation and less noise reduction compared to the expected trade-off set by the user through the parameters $c_i, i = 1, \cdots, m$.

## IV. Proposed Convex Approximation Method

The proposed method is a semi-definite convex relaxation (SDCR) of the optimization problem in Eq. (11). First, we review two important properties that will be useful for understanding the proposed optimization problem.

*Property 1:* Any quadratic expression can be expressed as [16]

$$\mathbf{q}^H \mathbf{Z} \mathbf{q} = \text{tr}\left(\mathbf{q}^H \mathbf{Z} \mathbf{q}\right) = \text{tr}\left(\mathbf{q}\mathbf{q}^H \mathbf{Z}\right). \tag{12}$$

*Property 2:* We have the following equivalence relation [17]

$$\mathbf{Z} = \begin{bmatrix}\mathbf{A} & \mathbf{B} \\ \mathbf{B}^H & \mathbf{C}\end{bmatrix} \succeq 0 \Leftrightarrow$$

$$\mathbf{A} \succeq 0, \quad \left(\mathbf{I} - \mathbf{A}\mathbf{A}^\dagger\right) \mathbf{B} = \mathbf{0}, \quad \mathbf{S}_1 \succeq 0, \tag{13}$$

$$\mathbf{C} \succeq 0, \quad \left(\mathbf{I} - \mathbf{C}\mathbf{C}^\dagger\right) \mathbf{B}^H = \mathbf{0}, \quad \mathbf{S}_2 \succeq 0, \tag{14}$$

with $\mathbf{S}_1 = \mathbf{C} - \mathbf{B}^H \mathbf{A}^\dagger \mathbf{B}$ the generalized Schur complement of $\mathbf{A}$ in $\mathbf{Z}$, $\mathbf{S}_2 = \mathbf{A} - \mathbf{B}\mathbf{C}^\dagger \mathbf{B}^H$ the generalized Schur complement of $\mathbf{C}$ in $\mathbf{Z}$, and $\mathbf{A}^\dagger$ is the pseudo-inverse of $\mathbf{A}$ [18].

Before, we present the proposed convex optimization problem, we first introduce an equivalent optimization problem to the problem in Eq. (11). That is,

$$\begin{aligned}\hat{\mathbf{w}}_L, \hat{\mathbf{w}}_R = \arg\min_{\mathbf{w}_L,\mathbf{w}_R} & \begin{bmatrix}\mathbf{w}_L^H & \mathbf{w}_R^H\end{bmatrix} \tilde{\mathbf{P}} \begin{bmatrix}\mathbf{w}_L \\ \mathbf{w}_R\end{bmatrix} \\ \text{s.t.} \quad & \mathbf{w}_L^H \mathbf{a} = a_L^H \quad \mathbf{w}_R^H \mathbf{a} = a_R^H, \\ & \left| \frac{\mathbf{w}_L^H \mathbf{b}_i}{\mathbf{w}_R^H \mathbf{b}_i} - \frac{b_{iL}}{b_{iR}} \right|^2 \leq \mathcal{E}_i^2, \; i = 1, \cdots, m \leq r.\end{aligned} \tag{15}$$

By reformulating the inequality in Eq. (15), we obtain an equivalent quadratic constraint given by

$$\begin{aligned}& \left| \frac{\mathbf{w}_L^H \mathbf{b}_i}{\mathbf{w}_R^H \mathbf{b}_i} - \frac{b_{iL}}{b_{iR}} \right|^2 \leq \mathcal{E}_i^2 \Rightarrow \\ & \underbrace{\begin{bmatrix}\mathbf{w}_L^H & \mathbf{w}_R^H\end{bmatrix}}_{\mathbf{w}^H} \underbrace{\begin{bmatrix}\mathbf{A} & \mathbf{B} \\ \mathbf{B}^H & \mathbf{C}\end{bmatrix}}_{\mathbf{M}_i} \underbrace{\begin{bmatrix}\mathbf{w}_L \\ \mathbf{w}_R\end{bmatrix}}_{\mathbf{w}} \leq 0,\end{aligned} \tag{16}$$

where $\mathbf{A} = |b_{iR}|^2\mathbf{b}_i\mathbf{b}_i^H$, $\mathbf{B} = -b_{iL}^* b_{iR}\mathbf{b}_i\mathbf{b}_i^H$, $\mathbf{C} = \left(|b_{iL}|^2 - |b_{iR}|^2\mathcal{E}_i^2\right)\mathbf{b}_i\mathbf{b}_i^H$. Therefore, the optimization problem in Eq. (15) can be re-written as

$$\begin{aligned}\hat{\mathbf{w}} = \arg\min_{\mathbf{w}} \ & \mathbf{w}^H\tilde{\mathbf{P}}\mathbf{w} \\ \text{s.t.} \quad & \mathbf{w}^H \begin{bmatrix}\mathbf{a} & \mathbf{0} \\ \mathbf{0} & \mathbf{a}\end{bmatrix} = \begin{bmatrix} a_L^* \\ a_R^* \end{bmatrix}, \\ & \mathbf{w}^H\mathbf{M}_i\mathbf{w} \leq 0, \quad i = 1, \cdots, m. \end{aligned} \tag{17}$$

The matrix $\mathbf{M}_i$ is not positive semi-definite and, therefore, the quadratic inequality constraint is not convex and, hence, the optimization problem in Eq. (17) is not convex. The proof of non positive semi-definiteness of $\mathbf{M}_i$ uses Property 2. Specifically, note that $\mathbf{A} \succeq 0$, but $\mathbf{S}_1 = -|b_{iR}|^2\mathcal{E}_i^2\mathbf{b}_i\mathbf{b}_i^H \preceq 0$, because $\mathbf{b}_i\mathbf{b}_i^H \succeq 0$ and $-|b_{iR}|^2\mathcal{E}_i^2 \leq 0$ and, therefore, $\mathbf{M}_i$ is not positive semi-definite.

The optimization problem in Eq. (17) is a non-convex quadratic-constrained quadratic program (QCQP) [17], [19]. Following the methodology described in [19], we use Property 1 to re-write the optimization problem in Eq. (17) into the following equivalent formulation:

$$\begin{aligned}\hat{\mathbf{w}}, \hat{\mathbf{W}} = \arg\min_{\mathbf{w},\mathbf{W}} \ & \operatorname{tr}\left(\mathbf{W}\tilde{\mathbf{P}}\right) \\ \text{s.t.} \quad & \mathbf{w}^H \begin{bmatrix}\mathbf{a} & \mathbf{0} \\ \mathbf{0} & \mathbf{a}\end{bmatrix} = \begin{bmatrix} a_L^* \\ a_R^* \end{bmatrix}, \\ & \operatorname{tr}\left(\mathbf{W}\mathbf{M}_i\right) \leq 0, \quad i = 1, \cdots, m, \\ & \mathbf{W} = \mathbf{w}\mathbf{w}^H. \end{aligned} \tag{18}$$

The optimization problem in Eq. (18) is still not convex, but it has two differences with the problem in Eq. (17). The trace inequality is convex, but the new equality constraint, $\mathbf{W} = \mathbf{w}\mathbf{w}^H$ is not convex. Following [19], we apply the SDCR to the non-convex equality constraint of the problem in Eq. (18) and obtain the convex optimization problem given by

$$\begin{aligned}\hat{\mathbf{w}}, \hat{\mathbf{W}} = \arg\min_{\mathbf{w},\mathbf{W}} \ & \operatorname{tr}\left(\mathbf{W}\tilde{\mathbf{P}}\right) \\ \text{s.t.} \quad & \mathbf{w}^H \begin{bmatrix}\mathbf{a} & \mathbf{0} \\ \mathbf{0} & \mathbf{a}\end{bmatrix} = \begin{bmatrix} a_L^* \\ a_R^* \end{bmatrix}, \\ & \operatorname{tr}\left(\mathbf{W}\mathbf{M}_i\right) \leq 0, \quad i = 1, \cdots, m. \\ & \mathbf{W} \succeq \mathbf{w}\mathbf{w}^H. \end{aligned} \tag{19}$$

Using Property 2, the inequality constraint $\mathbf{W} \succeq \mathbf{w}\mathbf{w}^H$ can be re-written as a linear matrix inequality, and the optimization problem in Eq. (19) can be re-written into a standard-form semi-definite program [19]. That is,

$$\begin{aligned}\hat{\mathbf{w}}, \hat{\mathbf{W}} = \arg\min_{\mathbf{w},\mathbf{W}} \ & \operatorname{tr}\left(\mathbf{W}\tilde{\mathbf{P}}\right) \\ \text{s.t.} \quad & \mathbf{w}^H \begin{bmatrix}\mathbf{a} & \mathbf{0} \\ \mathbf{0} & \mathbf{a}\end{bmatrix} = \begin{bmatrix} a_L^* \\ a_R^* \end{bmatrix}, \\ & \operatorname{tr}\left(\mathbf{W}\mathbf{M}_i\right) \leq 0, \quad i = 1, \cdots, m. \\ & \begin{bmatrix}\mathbf{W} & \mathbf{w} \\ \mathbf{w}^H & 1\end{bmatrix} \succeq 0. \end{aligned} \tag{20}$$

This is a convex optimization problem, which can be solved efficiently [19]. If the solutions are on the boundary, i.e., $\hat{\mathbf{W}} = \hat{\mathbf{w}}\hat{\mathbf{w}}^H$, the minimizer, $\hat{\mathbf{w}}$, of the problem in Eq. (20) is also the minimizer of the non-convex RBB problem. This means, that in the case of $\hat{\mathbf{W}} = \hat{\mathbf{w}}\hat{\mathbf{w}}^H$, the proposed problem in Eq. (20) is optimal. Moreover, in this case, the inequalities of the problem in Eqs. (17), (15) (11) are satisfied. Otherwise, if $\hat{\mathbf{W}} \succ \hat{\mathbf{w}}\hat{\mathbf{w}}^H$, the solution of the problem in Eq. (20) may or may not satisfy the inequalities of the RBB, which means that we lose the guarantee for user-controlled upper-bounded ITF error when the (R)ATF vectors of the interferers have been estimated accurately enough. In our experience, in practice $\hat{\mathbf{W}} = \hat{\mathbf{w}}\hat{\mathbf{w}}^H$ almost never happens. Nevertheless, we will experimentally show in Section V that the SDCR method always stays relatively close to the boundary of the inequality constraints of the RBB problem. Finally, the main advantage of the new proposed SDCR method is that it reduces significantly the computational complexity, since a single convex optimization problem is solved compared to the multiple convex optimization problems that must be solved in the SCO method.

### *A. Proposed Hybrid Method*

In this section, we propose a hybrid method, which combines the SDCR and the SCO methods into a single method. If the (R)ATF vectors of the interferers are estimated accurately enough, the hybrid method guarantees user-controlled upper-bounded binaural-cue distortions of the interferers as in the first version of the SCO method. Moreover, the proposed hybrid method is significantly faster than the SCO method and slightly slower than the SDCR method. We will experimentally show in Section V, that the hybrid proposed method achieves solutions closer to the boundary of the inequality constraints of the RBB problem compared to the SCO method, while at the same time achieving more noise suppression.

For a particular frequency bin, the hybrid method first solves the SDCR problem and then checks if the inequality constraints of Eq. (11) are satisfied. If all of them are satisfied, the SDCR method will be used to approximately solve the RBB problem. Otherwise the SCO method is used to approximately solve the RBB problem in this particular frequency bin. In such a way, there is a guarantee that we will always have an optimal solution which satisfies the constraints of the RBB problem, while at the same time reducing the overall computational complexity significantly. In order to avoid switching to the SCO method for just negligibly larger ITF errors than the user-controlled upper bounds $\mathcal{E}_i$, we use the following switching criterion:

$$\left| \frac{\mathbf{w}_L^H\mathbf{b}_i}{\mathbf{w}_R^H\mathbf{b}_i} - \frac{b_{iL}}{b_{iR}} \right| \leq \tilde{\mathcal{E}}_i, \ i = 1, \cdots, m, \tag{21}$$

where $\tilde{\mathcal{E}}_i$ is a slightly increased upper bound and is given by

$$\tilde{\mathcal{E}}_i = (c_i + \epsilon)\left| \frac{a_L}{a_R} - \frac{b_{iL}}{b_{iR}} \right|, \ i = 1, \cdots, m, \tag{22}$$

where $\epsilon$ is very small, e.g., $0 < \epsilon < 0.1$. This modification avoids possible switching to the SCO method for negligibly larger ITF errors than the $\mathcal{E}_i$. The hybrid method is summarized in Algorithm 1.

**Algorithm 1:** Hybrid scheme

$\hat{\mathbf{w}}_1 \leftarrow$ SDCR Problem in Eq. (20)
**if** $\hat{\mathbf{w}}_1$ satisfies Eq. (21) **then**
  return $\hat{\mathbf{w}}_1$
**else**
  $\hat{\mathbf{w}}_2 \leftarrow$ SCO method [12]
  return $\hat{\mathbf{w}}_2$
**end if**

## V. Experiments

We conducted two different sets of experiments: the first examines the performance difference between the SCO method [12] (with $k_{\max} = 50$), the proposed SDCR method, and the proposed hybrid (with $\epsilon = 0.05$) method, when the true RATF vectors of the interferers are used. The reason for that is to show the theoretical trade-off between noise reduction and binaural-cue preservation. The second experiment examines the performance of the same methods, when the pre-determined RATF vectors are used for preserving the binaural cues of the interferers. Note that in both sets of experiments, we used the true RATF vector of the target source. We used the CVX toolbox [20] to solve the convex optimization problems associated with the SCO, SDCR and hybrid methods. The CVX toolbox uses an interior point method to solve the convex optimization problems [17]. In all methods that approximately solve the RBB problem, we used a common $c$ value for all interferers in the inequality constraints, i.e., $c_i = c, \forall i$. We also included the BMVDR BF as a reference method in the comparisons. The noise cross-power spectral density matrix was estimated using 5 seconds of a noise-only segment, where all interferers are active, but the target source is inactive. The spatial filters of all methods were estimated only once using the same estimated noise cross-power spectral density matrix and, thus, they are time invariant.

Note that for the pre-determined RATF vectors, we used the RATF vectors of 24 pre-determined anechoic head impulse responses from the database in [21]. The pre-determined RATF vectors are associated with azimuths uniformly spaced around the head with a resolution of $360/24 = 15$ degrees, starting from $-90$ degrees. Please note that the pre-determined RATF vector at 0 degrees was omitted from the constraints, because it was in the same direction as the RATF vector of the target.

### A. Acoustic Scene Setup

The acoustic scene that we used consists of one target female talker in the look direction (i.e., 0 degrees), and 4 interferers, where each has the same average power at its original location, as the target signal at the original location. The first interferer is a male talker on the right-hand side of the HA user with azimuth of 80 degrees; the second interferer is a music signal on the right-hand side of the HA user with azimuth of 50 degrees; the third interferer is a vacuum cleaner on the left-hand side of the HA user with azimuth $-35$ degrees; and the fourth interferer is a ringing mobile phone on the left-hand side with azimuth $-70$ degrees. Note that the RATF vectors of all interferers have an azimuth mismatch with the pre-determined RATF vectors' azimuths. The microphone self-noise is set to have a 40 dB SNR at the left reference microphone, and it has the same power in all microphones.

### B. Hearing-Aid Setup and Processing

The total number of microphones is $M = 4$; two at each HA. The sampling frequency is 16 kHz. The microphone signals were constructed using the head impulse responses from the reverberant office environment from the database in [21]. We used the overlap-and-add processing method [22] for analyzing and synthesizing our signals. The analysis and synthesis windows are square-root Hanning windows and the overlap is 50%. The frame length is 10 ms, i.e., 160 samples, and the FFT size is 256.

### C. Evaluation Methodology

We measure the noise-reduction performance in terms of the segmental signal-to-noise-ratio (SSNR) only in target-presence time regions. We used an ideal activity detector to find these time-regions. We also predict intelligibility with the STOI measure [23].

We measure binaural-cue distortions with instrumental measures and a listening test. The instrumental measures are the average ITF error, interaural level difference (ILD) error and interaural phase difference (IPD) error per interferer. These averages are calculated only over frequency, since we have fixed BFs over time. Note that, for the IPD error, we averaged only the frequency bins in the range of $0 - 1.5$ kHz, while for the ILD error, we averaged only the frequency bins in the range of $3 - 8$ kHz. The reason for this choice is that the ILDs are perceptually more important for localization above 3 kHz, while the IPDs are perceptually more important for localization below 1.5 kHz [24]. Note that we used the expressions from [13] for computing the ILD and IPD errors for a single frequency bin. We do not measure the binaural-cue distortions of the target, because all methods achieve perfect preservation of the binaural-cues of the target, since i) there are no estimation errors on the RATF vector of the target signal used in the associated optimization problems and ii) the response of the binaural spatial filter with respect to the target at the two reference microphones is distortionless.

The listening test is performed using the methodology described in [6], and examines the performance of the compared methods only in the case of the pre-determined RATF vectors. Ten normal-hearing subjects participated, excluding the authors. They were asked to determine the azimuths of all point-sources in the acoustic scene when listening to signals processed by the compared methods as well as the unprocessed scene. The tested $c$ values were 0.3 and 0.7 for the SCO, SDCR and hybrid methods. In addition to listening to the noisy and processed signals, the subjects also listened to the clean unprocessed point sources in isolation, in order to determine the reference azimuthms of the point sources. The localization errors were calculated with respect to the reference (and not the true) azimuths as in [6]. This is because we used only one set of head impulse responses from [21] to construct the binaural signals, which means that every subject will have a different reference azimuth. In this way, a significant estimation bias was removed. Two repetitions of the listening test were conducted. The reference azimuth of each source and every subject was computed as the average between the two repetitions, and the error was computed with respect to this averaged reference azimuth. The localization errors of the sources were averaged over subjects and repetitions. A t-test was used in order determine whether the methods result in statistically significantly different perceived source locations.

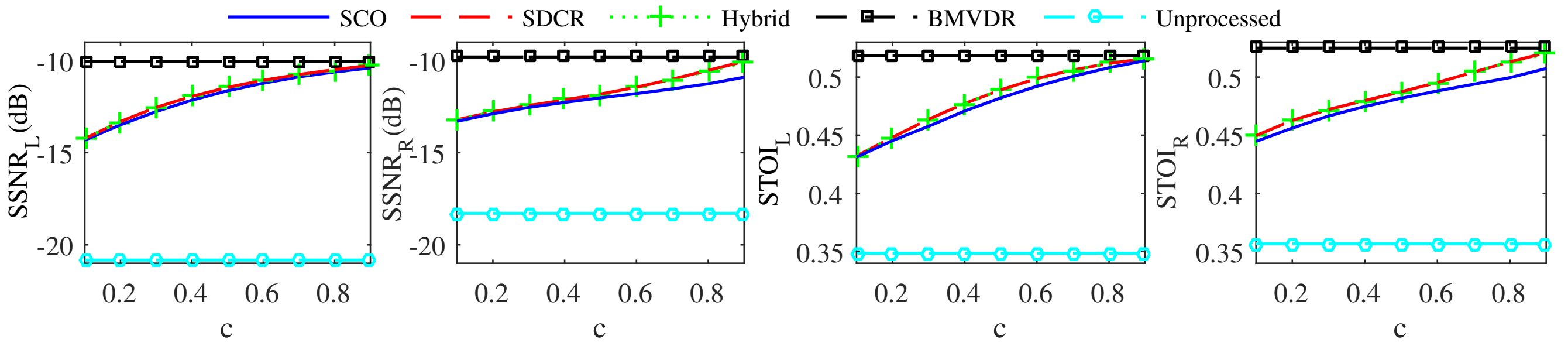

Fig. 1: Noise reduction and intelligiblity prediction performances when the true RATF vectors of the interferers are used in the SCO, SDCR and hybrid methods.

We also measured the complexity of the compared methods in terms of the number of convex optimization problems that they needed to solve for all frequency bins in total. Note that the BFs are fixed over time and, therefore, we do not measure varying complexity over time.

### *D. Discussion of Results with True RATF Vectors*

In this section, the compared methods use the true RATF vectors of the sources in the constraints. Fig. 1 depicts the noise reduction performance and intelligibility prediction of the unprocessed scene, and SCO, SDCR, BMVDR methods at both reference microphones. As expected the BMVDR achieves the best noise reduction performance and predicted intelligibility. It is clear, that all other methods achieve similar performances for the left reference microphone, while for the right reference microphone the SCO method achieves the worst noise reduction performance among all. Moreover, as expected, as $c$ increases, the noise reduction and STOI value increases for all methods. Note that the SDCR method has almost identical performance as the hybrid method. This is because, in this example the hybrid method switched to the SCO method only a few times.

Fig. 2 shows the binaural-cue distortions of the compared methods per interfering source. As expected, the larger binaural-cue distortions are obtained with the BMVDR BF, while all other methods achieve less binaural-cue distortions. As expected, as $c$ increases, the binaural-cue distortions increase. Note that for the ITF errors, we also display the $c$ times the ITF error of the BMVDR (which is labeled as ITF upper bound) in order to visualize the closeness of the estimated spatial filters at the boundary of the inequality constraints of the RBB problem. It is clear that both SDCR and hybrid methods are closer to the boundary of the inequality constraints compared to the SCO method. Moreover, the hybrid method is for all $c$ values (on average) below the boundary, even if we used the extended switch criterion in Eq. (21). On the other hand, the ITF error of the SDCR method sometimes (see Interferers 1 and 2) is slightly above the boundary. As explained in Section IV, this is because the SDCR method does not guarantee a user-controlled upper-bounded ITF error as the SCO or the hybrid methods do. Note also that as expected the SCO method for e.g., $c = 0.8, 0.9$ values, is not close to the boundary, while the SDCR and hybrid methods are closer to the boundary.

Fig. 3 shows the computational complexity of the compared methods in terms of number of convex optimization problems required to solve for convergence. The SDCR method requires to solve much less convex problems than the SCO method (especially at larger $c$ values) and slightly less compared to the hybrid method. The hybrid method again requires to solve much less convex problems than the SCO method, especially at larger $c$ values.

We can conclude from the above that, in most cases, the theoretical performance (i.e., when the true RATF vectors are used) of both proposed methods is more optimal than the SCO method. Specifically, both proposed methods provide solutions that are closer to the expected solutions of the RBB problem, since both proposed methods are closer to the boundary. This means that both methods provide a more user-controlled trade-off between noise reduction and binaural-cue preservation than the SCO method, especially in large $c$ values. Finally both proposed methods are significantly less computationally demanding than the SCO method.

### *E. Discussion of Results with Pre-Determined RATF Vectors*

In this section, the compared methods use the pre-determined RATF vectors. Fig. 4 shows the noise reduction performance and intelligibility prediction of the compared methods. Here the gap in performance between the proposed methods and the SCO method is bigger compared to the case where the true RATF vectors were used. The proposed methods (especially the SDCR method) significantly improved both noise reduction and predicted intelligibility at both reference microphones. The reason why the performance gap between the SDCR method and the hybrid method is increased compared to the case where the true RATF vectors were used is because the hybrid method switched many more times to the SCO method (see Algorithm 1) here. In conclusion, both proposed methods achieve in most cases a better noise reduction and predicted intelligibility than the SCO method, especially for larger $c$ values.

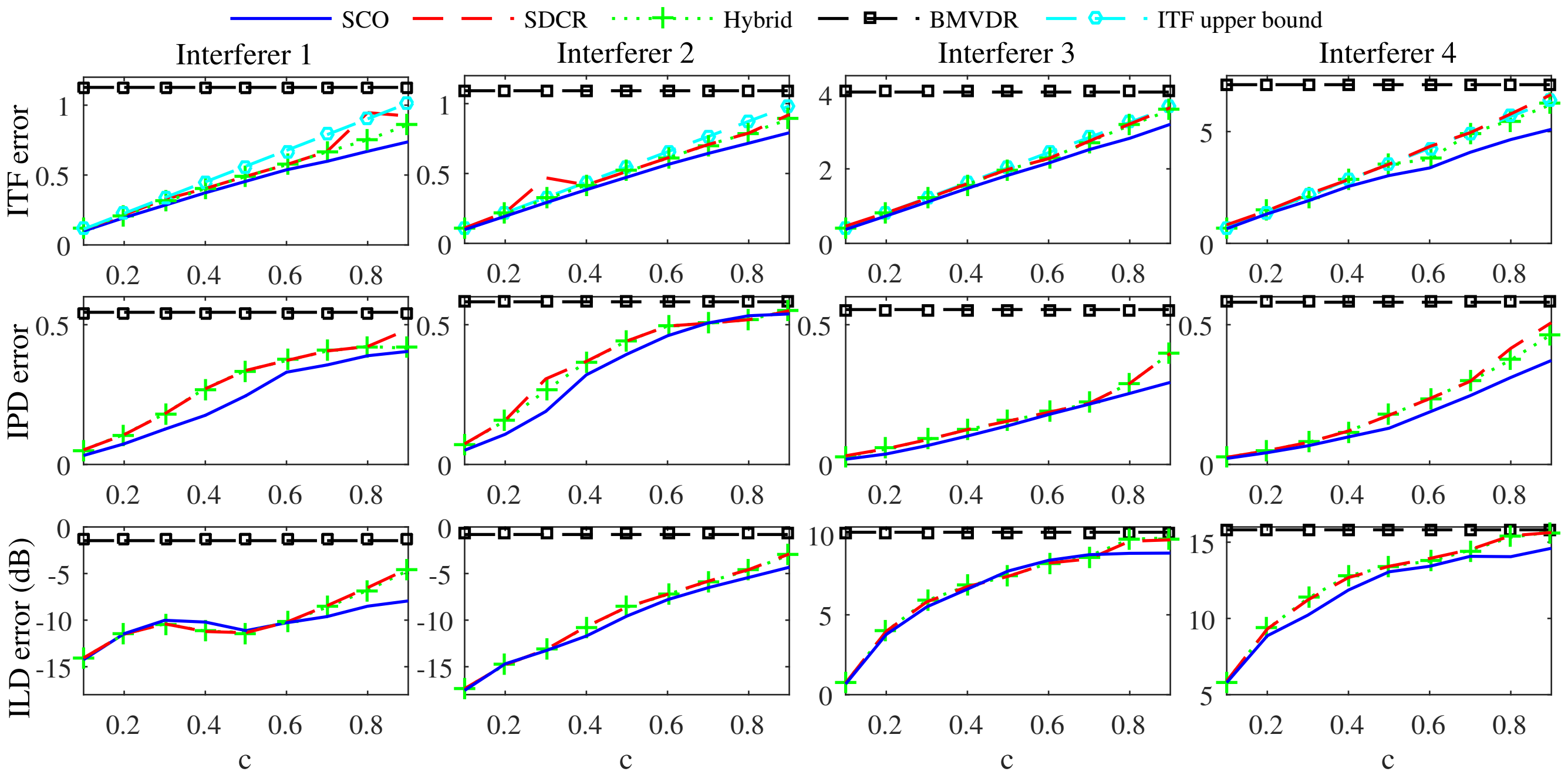

Fig. 2: Binaural-cue distortions (averaged over frequency) of interferers when the true RATF vectors of the interferers are used in the SCO, SDCR and hybrid methods.

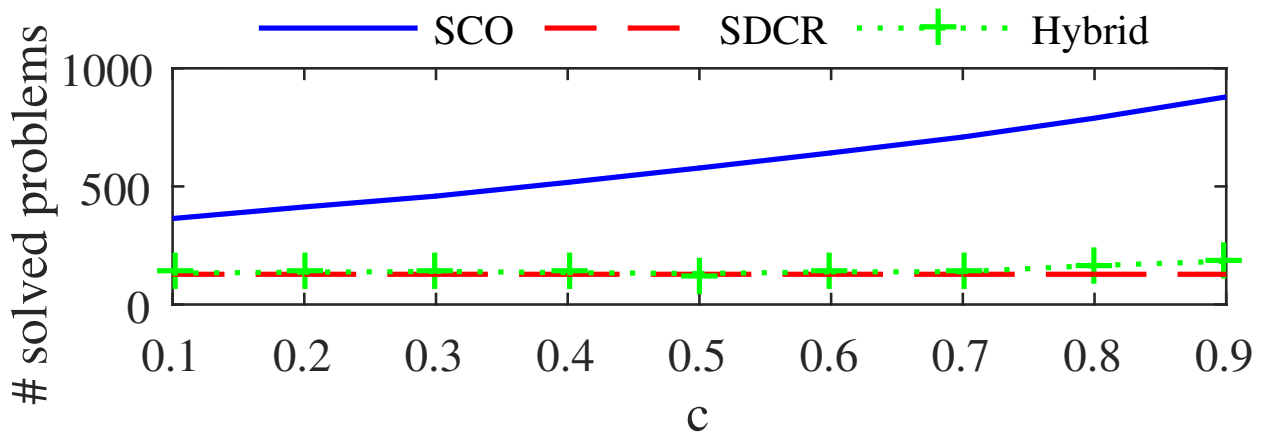

Fig. 3: Computational complexity measured as the number of solved convex optimization problems (in all frequency bins) when the true RATF vectors of the interferers are used in the SCO, SDCR and hybrid methods.

Fig. 5 shows the binaural-cue distortions of the compared methods per interfering source. As expected, when predetermined RATF vectors are used, all methods do not guarantee user-controlled upper-bounded ITF error of the interferers. Therefore, all methods, in many occasions (see interferers 3 and 4), result in a larger ITF error than the average ITF upper bound of the RBB problem when computed using the true RATFs of the interferers. The SCO method has the lowest binaural-cue distortions compared to the compared methods. Nevertheless, we will see later on in the t-test of the listening test that the compared methods are not significantly different for the same $c$ values.

In Fig. 6, we show the computational complexities of the compared methods. Again the SDCR method requires to solve less convex problems compared to the SCO method, but the hybrid method does not have a huge computational advantage over the SCO method in this case. However, the usage of the hybrid method using pre-determined (R)ATF vectors is not critical, since anyway no method can guarantee user-controlled upper-bounded ITF error of the interferers, unless the number of pre-determined RATF vectors is huge. This of course is not practical since it may result in non-feasible solutions and/or the noise reduction will be negligible due to the large number of constraints.

Fig. 7 shows the results of the subjective localization test. A similar behavior as with the instrumental binaural-cue distortion measures is observed here. The only difference appears for the ringing mobile phone, where for $c = 0.7$ all methods achieve slightly worse performance than the BMVDR. Several users also reported difficulty in localizing the ringing phone after completing the test. We believe that this is because of the high frequency content of the ringing tone of the mobile phone and only the ILDs might have been used for localization.

Table I shows the results of the t-test, which was done by gathering all localization errors of all sources. The significance level was set to $5\%$. It is clear that the SCO, SDCR and hybrid methods are all not significantly different for the same $c$ value. This means that even though we observed less binaural-cue distortions in the SCO method in Figs 2 and 5, compared to the proposed methods for the same $c$ value, these differences are not perceptually important. However, recall that the proposed methods achieve a better noise reduction and predicted intelligibility compared to the SCO method. Thus, the proposed methods provide a better perceptual trade-off compared to the SCO method. Finally, note that the SCO, SDCR and hybrid methods are not statistically significantly different from the unprocessed scene for $c = 0.3$. This means that in all three methods the subjects managed (on average) to localize as good as in the unprocessed scene. However, unlike the unprocessed scene, all three methods improved noise reduction and predicted intelligibility.

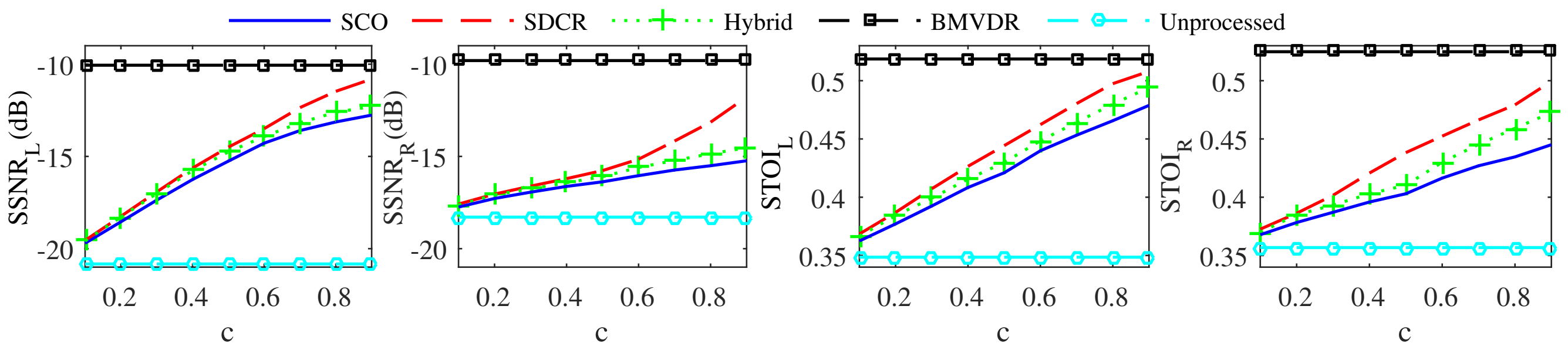

Fig. 4: Noise reduction and intelligiblity prediction performances when the pre-determined RATF vectors of the interferers are used in the SCO, SDCR and hybrid methods.

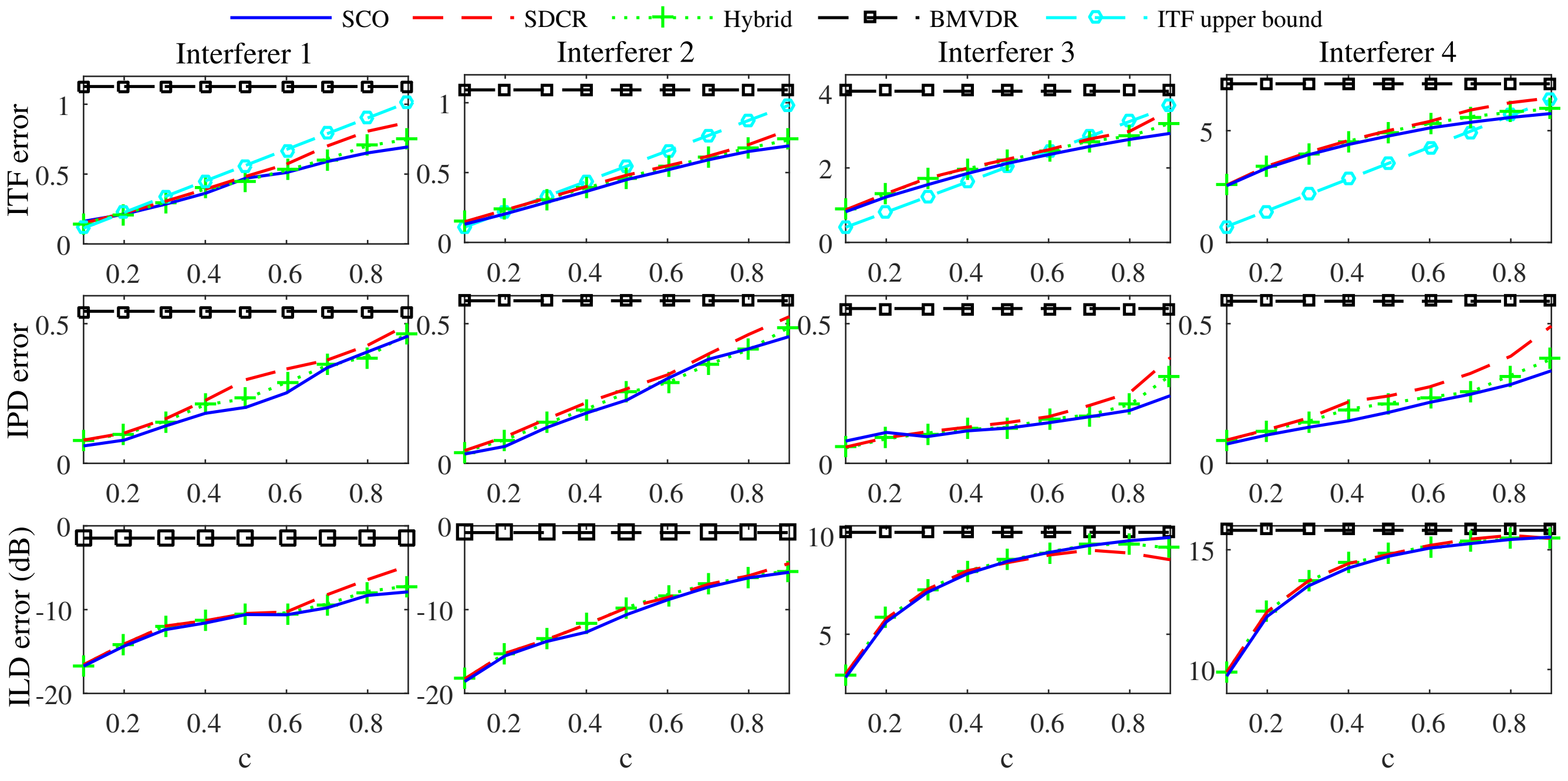

Fig. 5: Binaural-cue distortions (averaged over frequency) of interferers when the pre-determined RATF vectors of the interferers are used in the SCO, SDCR and hybrid methods.

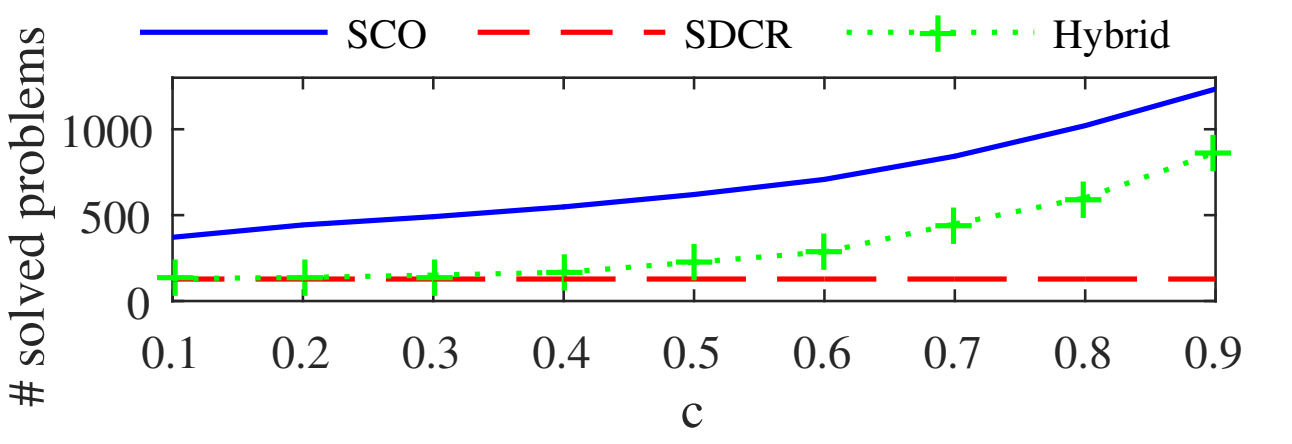

Fig. 6: Computational complexity measured as the number of solved convex optimization problems (in all frequency bins) when the pre-determined RATF vectors of the interferers are used in the SCO, SDCR and hybrid methods.

## VI. Conclusion

We proposed two new suboptimal methods for approximately solving the non-convex relaxed binaural beamforming (RBB) optimization problem. Both methods are significantly computationally less demanding compared to the existing successive convex optimization (SCO) method. For each fre-quency bin, the SCO method requires to solve multiple convex optimization problems in order to converge. In contrast, the first proposed method, which is a semi-definite convex relaxation (SDCR) of the RBB problem, solves only one convex optimization problem per frequency bin. Apart from the computational advantage, the SDCR method also achieves in most cases a better trade-off between intelligibility and binaural-cue preservation than the SCO method. However, the SDCR method does not guarantee user-controlled upper bounded ITF error when the RATF vectors of the interferers are estimated accurately enough. This problem is solved by the second proposed method, which is a hybrid combination of the SDCR and SCO methods. This method guarantees user-controlled upper-bounded ITF error, and at the same time is computationally much less demanding than the SCO method. Finally, listening tests showed that all three methods achieve the same localization errors for the same amount of relaxation.

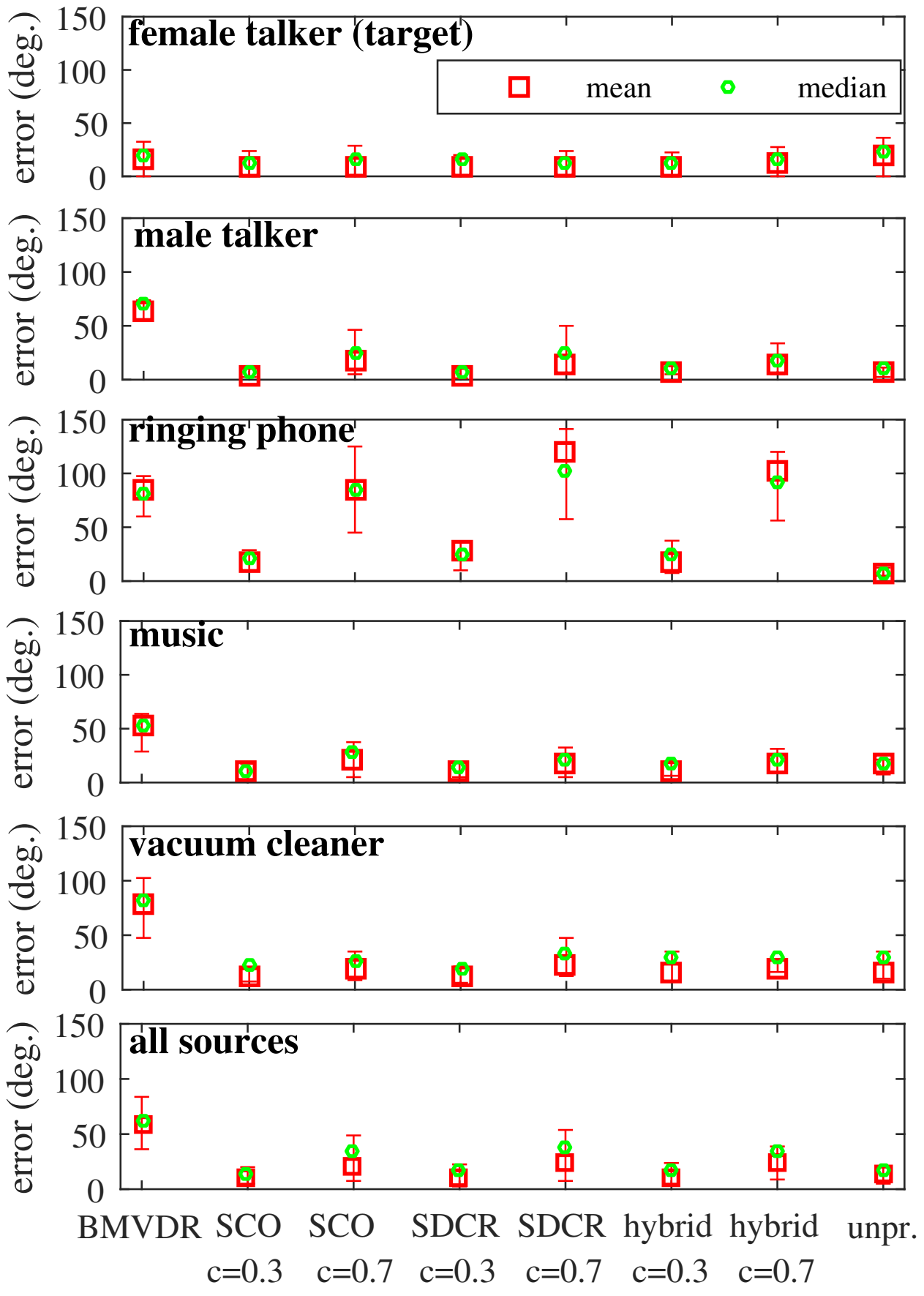

Fig. 7: Localization test comparing the SCO, SDCR and hybrid methods with respect to the localization error in degrees.

TABLE I: T-test: + denotes significantly different (i.e., the null hypothesis is rejected at 5% significance level), while ∘ denotes not significantly different.

| Method | BMVDR | SCO $c=0.3$ | SCO $c=0.7$ | SDCR $c=0.3$ | SDCR $c=0.7$ | Hybrid $c=0.3$ | Hybrid $c=0.7$ |
|---|---|---|---|---|---|---|---|
| BMVDR | ∘ | + | + | + | + | + | + |
| SCO $c=0.3$ | + | ∘ | + | ∘ | + | ∘ | + |
| SCO $c=0.7$ | + | + | ∘ | + | ∘ | + | ∘ |
| SDCR $c=0.3$ | + | ∘ | + | ∘ | + | ∘ | + |
| SDCR $c=0.7$ | + | + | ∘ | + | ∘ | + | ∘ |
| Hybrid $c=0.3$ | + | ∘ | + | ∘ | + | ∘ | + |
| Hybrid $c=0.7$ | + | + | ∘ | + | ∘ | + | ∘ |
| Unpro-cessed | + | ∘ | + | ∘ | + | ∘ | + |

## References

[1] S. Doclo, W. Kellermann, S. Makino, and S. Nordholm, "Multichannel signal enhancement algorithms for assisted listening devices," *IEEE Signal Process. Mag.*, vol. 32, no. 2, pp. 18–30, Mar. 2015.

[2] J. M. Kates, *Digital hearing aids*. Plural publishing, 2008.

[3] E. Hadad, D. Marquardt, S. Doclo, and S. Gannot, "Theoretical analysis of binaural transfer function MVDR beamformers with interference cue preservation constraints," *IEEE Trans. Audio, Speech, Language Process.*, vol. 23, no. 12, pp. 2449–2464, Dec. 2015.

[4] A. W. Bronkhorst, "The cocktail party phenomenon: A review of research on speech intelligibility in multiple-talker conditions," *Acta Acoustica*, vol. 86, no. 1, pp. 117–128, 2000.

[5] D. Marquardt, "Development and evaluation of psychoacoustically motivated binaural noise reduction and cue preservation techniques," Ph.D. dissertation, Carl von Ossietzky Universität Oldenburg, 2015.

[6] A. I. Koutrouvelis, R. C. Hendriks, R. Heusdens, S. van de Par, J. Jensen, and M. Guo, "Evaluation of binaural noise reduction methods in terms of intelligibility and perceived localization," in *submitted to EUSIPCO*, 2018.

[7] J. G. Desloge, W. M. Rabinowitz, and P. M. Zurek, "Microphone-array hearing aids with binaural output .I. Fixed-processing systems," *IEEE Trans. Speech Audio Process.*, vol. 5, no. 6, pp. 529–542, Nov. 1997.

[8] D. P. Welker, J. E. Greenberg, J. G. Desloge, and P. M. Zurek, "Microphone-array hearing aids with binaural output .II. A two-microphone adaptive system," *IEEE Trans. Speech Audio Process.*, vol. 5, no. 6, pp. 543–551, Nov. 1997.

[9] T. Klasen, T. Van den Bogaert, M. Moonen, and J. Wouters, "Binaural noise reduction algorithms for hearing aids that preserve interaural time delay cues," *IEEE Trans. Signal Process.*, vol. 55, no. 4, pp. 1579–1585, Apr. 2007.

[10] A. I. Koutrouvelis, R. C. Hendriks, J. Jensen, and R. Heusdens, "Improved multi-microphone noise reduction preserving binaural cues," in *IEEE Int. Conf. Acoust., Speech, Signal Process. (ICASSP)*, Mar. 2016.

[11] E. Hadad, S. Doclo, and S. Gannot, "The binaural LCMV beamformer and its performance analysis," *IEEE Trans. Audio, Speech, Language Process.*, vol. 24, no. 3, pp. 543–558, Jan. 2016.

[12] A. I. Koutrouvelis, R. C. Hendriks, R. Heusdens, and J. Jensen, "Relaxed binaural LCMV beamforming," *IEEE Trans. Audio, Speech, Language Process.*, vol. 25, no. 1, pp. 137–152, Jan. 2017.

[13] B. Cornelis, S. Doclo, T. Van den Bogaert, M. Moonen, and J. Wouters, "Theoretical analysis of binaural multimicrophone noise reduction techniques," *IEEE Trans. Audio, Speech, Language Process.*, vol. 18, no. 2, pp. 342–355, Feb. 2010.

[14] S. Gannot, E. Vincet, S. Markovich-Golan, and A. Ozerov, "A consolidated perspective on multi-microphone speech enhancement and source separation," *IEEE Trans. Audio, Speech, Language Process.*, vol. 25, no. 4, pp. 692–730, April 2017.

[15] A. I. Koutrouvelis, R. C. Hendriks, R. Heusdens, J. Jensen, and M. Guo, "Binaural beamforming using pre-determined relative acoustic transfer functions," in *EURASIP Europ. Signal Process. Conf. (EUSIPCO)*, Aug. 2017.

[16] H. Anton, *Elementary linear algebra*. John Wiley & Sons, 2010.

[17] S. Boyd and L. Vandenberghe, *Convex optimization*. Cambridge university press, 2004.

[18] G. Golub and C. V. Loan, *Matrix Computations*, 3rd ed. Oxford: North Oxford Academic, 1983.
[19] L. Vandenberghe and S. Boyd, "Semidefinite programming," *SIAM review*, vol. 38, no. 1, pp. 49–95, Mar. 1996.
[20] "Cvx: Matlab software for disciplined convex programming." 2008.
[21] H. Kayser, S. Ewert, J. Annemuller, T. Rohdenburg, V. Hohmann, and B. Kollmeier, "Database of multichannel in-ear and behind-the-ear head-related and binaural room impulse responses," *EURASIP J. Advances Signal Process.*, vol. 2009, pp. 1–10, Dec. 2009.
[22] J. B. Allen, "Short-term spectral analysis, and modification by discrete Fourier transform," *IEEE Trans. Acoust., Speech, Signal Process.*, vol. 25, no. 3, pp. 235–238, June 1977.
[23] C. H. Taal, R. C. Hendriks, R. Heusdens, and J. Jensen, "An algorithm for intelligibility prediction of time-frequency weighted noisy speech," *IEEE Trans. Audio, Speech, Language Process.*, vol. 19, no. 7, pp. 2125–2136, Sep. 2011.
[24] W. M. Hartmann, "How we localize sound," *Physics Today*, vol. 52, no. 11, pp. 24–29, Nov. 1999.